\documentclass[aps,amsmath,amssymb,superscriptaddress,twocolumn,10pt,prl,floatfix,showpacs]{revtex4} 
\usepackage{graphicx}
\usepackage{psfrag}
\DeclareMathOperator{\Tr}{Tr}
\begin{document}

\title{Superfluidity and magnetism in multicomponent ultracold fermions}
\author{R. W. Cherng}
\affiliation{
Department of Physics, 
Harvard University, 
Cambridge, Massachusetts 02138, USA}
\author{G. Refael}
\affiliation{
Department of Physics, 
California Institute of Technology,
 Pasadena, California 91125, USA}
\author{E. Demler}
\affiliation{
Department of Physics, 
Harvard University, 
Cambridge, Massachusetts 02138, USA}
\date{\today}
\begin{abstract}
We study the interplay between superfluidity and magnetism in a multicomponent gas of
ultracold fermions.  Ward-Takahashi identities constrain
possible mean-field states describing order parameters for both pairing and
magnetization.  The structure of global phase
diagrams arises from competition among these states as functions of anisotropies
in chemical potential, density, or interactions.  They exhibit first and second
order phase transition as well as multicritical points, metastability regions,
and phase separation.  We comment on experimental signatures in ultracold atoms.

\end{abstract}
\date{\today}
\maketitle

Fermionic $s$-wave superfluidity
requires pairing between fermions with different 
internal states.
However, the nature of superfluidity for $N=2$ component 
systems and $N\ge 3$ is fundamentally different.  Superfluidity
\textit{is suppressed by} magnetization for $N=2$ because
there is one way to pair and not every particle can find a partner.
This led to proposals for exotic paired
states with broken translational symmetry \cite{ff-64,lo-64}
or gapless excitations \cite{sarma-63,liu-03}.
In contrast, superfluidity \textit{drives} magnetization for $N\ge 3$.
Here condensation energy favors enhancing the population for paired
components to different degrees depending on the density of 
states and interaction energy.  This occurs by cannibalizing
the populations for unpaired components. 

Ultracold atoms offer direct access to multicomponent
fermionic superfluids.
Observation of two component, equal population superfluids 
used Feshbach resonances to tune interactions
\cite{regal-04,zwierlein-04,bartenstein-04,kinast-04}.
The population for each component is both tunable
and essentially conserved, enabling
later experiments on superfluidity with imbalance
\cite{mit-06,rice-06}.
Recently, scattering lengths and locations of overlapping
Feshbach resonances between all three of the nearly degenerate
lowest sublevels of $^6$Li were measured \cite{bartenstein-05}.
This suggests experiments with $N\ge 3$ are within reach.

\begin{figure}
\includegraphics[width=2in]{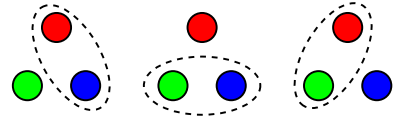}
\caption{(color online) Pairing (ellipses) two components (colors) in all possible ways generates
the diagonal pairing states as shown for $N=3$.  Only these states generically satisfy the constraints
on the microscopic pairing wavefunction imposed by Ward-Takahashi identities derived in the text, linear
combinations do not. Paired (unpaired) components have gapped (gapless) quasiparticle excitations.}
\label{fig:diagonal_pairing_states}
\end{figure}

In this Letter, we consider the interplay of superfluidity and magnetism
within a mean-field theory of $N\ge 3$ multicomponent fermions, each 
individually conserved.
We derive a Ward-Takahashi (WT) identity (Eq. \ref{eq:boson_wt_identity})
which provides fundamental constraints on possible mean-field states 
describing both pairing $\Delta$ and magnetization $M$.
We demonstrate this WT identity naturally leads to a specific form
for the microscopic pairing wavefunction which we call 
diagonal pairing states (DPS) illustrated in Fig. \ref{fig:diagonal_pairing_states}.
In these states, gapless excitations always exist for $N$ odd and are still 
possible for $N$ even.

\begin{figure}
\begin{tabular}[b]{cccc}
\includegraphics[width=1.5in]{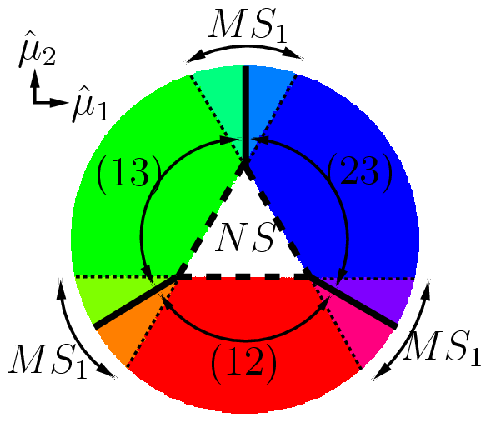}&\ \includegraphics[width=1.5in]{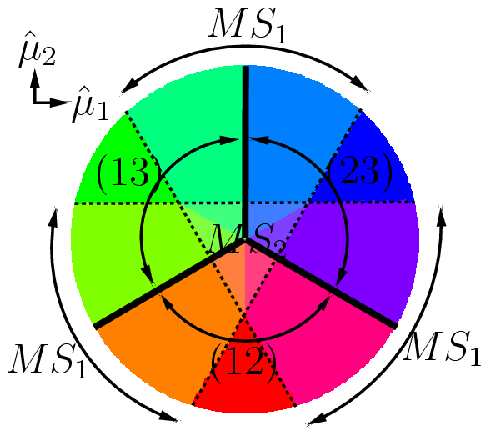}\\
\includegraphics[width=1.5in]{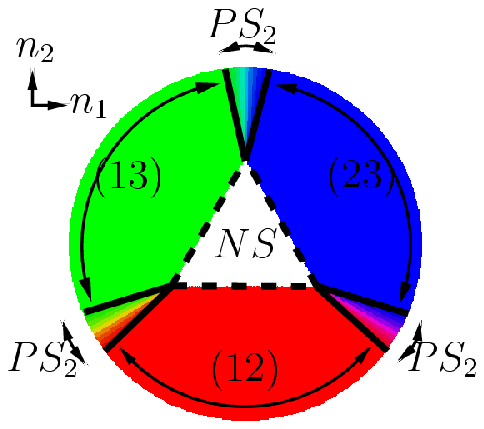}&\ \includegraphics[width=1.5in]{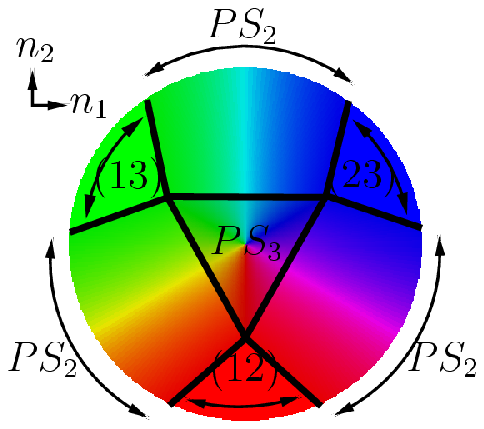}
\end{tabular}
\caption{(color online) $N=3$ phase diagrams for $T>T_c^{SYM}$ (left) and $T<T_c^{SYM}$ (right) against 
anisotropies in chemical potential $\hat{\mu}_\alpha$ (top) and density $n_\alpha$ (bottom).
$NS$ denotes the normal state and $(\alpha\beta)$ a state with $\alpha$, $\beta$
paired.  Solid (dashed) lines are first (second) order transitions while dotted lines
are boundaries for metastability regions.  $MS_i$ denotes regions with $i$ additional
metastable paired states and $PS_i$ a phase-separated mixture of $i$ paired states.}
\label{fig:n3_phasediagram}
\end{figure}

\begin{figure}
\includegraphics[width=3in]{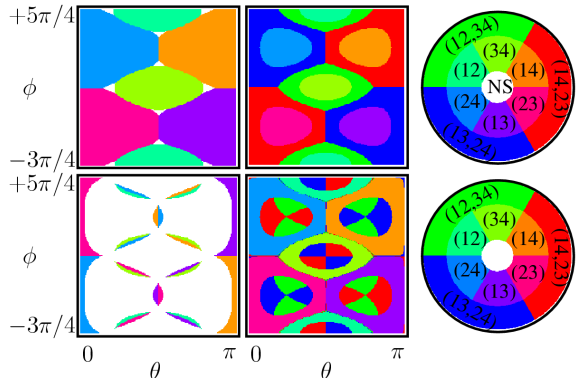}
\caption{(color online) Representative $N=4$ phase diagrams for $T>T_c^{SYM}$ (left) and $T<T_c^{SYM}$ (middle) against
0 in chemical potential $\mu_\alpha$ with legend (right) for the global minimum (top) and first metastable state (bottom).
$(\alpha_i\beta_i,\ldots)$ denotes $\alpha_i$, $\beta_i$ paired and white regions $NS$ the normal state
(top) or absence of metastable states (bottom).
First (second) order transitions separate different paired states 
(paired states from the normal state).}
\label{fig:n4_phasediagram}
\end{figure}

Having classified the mean-field states, we derive
global phase diagrams describing the competition among the
DPS as shown in Figs. \ref{fig:n3_phasediagram} and \ref{fig:n4_phasediagram}.
We focus on phase diagrams tuned by anisotropies in chemical potential or density
but the structure is the same for anisotropies in interactions.
For $N\ge 3$, $\Delta\Delta^\dagger$ acts as an external field for $M$
through $M\Delta\Delta^\dagger$.
This coupling vanishes identically on group theoretical grounds for $N=2$ where
$M$ couples to $\Delta$ only at higher order.
The structure of the DPS shows pairing \textit{always} drives magnetization for $N$ odd
and \textit{generically} does so for $N$ even through the coupling $M\Delta\Delta^\dagger$.
First order transitions and corresponding metastability as well as phase separation regions 
separate different DPS while second order transitions separate DPS and the normal state.
These transitions terminate at bicritical and multicritical points.

Previous theoretical works have also addressed superfluditiy with $N\ge 3$ components of fermions. 
This includes analysis of mean-field states for $N=3$ 
\cite{hofstetter-04a,hofstetter-04b} and $N=4$
\cite{wu-03} as well as phase diagrams for $N=3$ \cite{rapp-06,paananen-06}.
We focus on classification of allowed pairing states through
general symmetry arguments and Ward-Takahashi identities.
This complementary approach allows us to obtain generic and robust features
of both the resulting states and phase diagrams as well as providing a unified
perspective on pairing in multicomponent fermionic systems.

We consider the action $S(H)=S_0+S_{int}(\Gamma)$ where
\begin{equation}
\begin{split}
S_0(H)&=\sum_{\alpha\beta}
\left[\left(\partial_\tau+\frac{\nabla^2}{2m}\right)\delta_{\alpha\beta}-H_{\alpha\beta}\right]\bar{\psi}_\alpha \psi_\beta,\\
S_{int}(\Gamma)&=-\sum_{\alpha\beta\gamma\delta}
\Gamma_{\alpha\beta\gamma\delta}\bar{\psi}_{\alpha}\bar{\psi}_{\beta}\psi_{\gamma}\psi_{\delta}
\end{split}
\end{equation}
describing a dilute gas of fermions with mass $m$ and attractive contact 
interactions where
$\psi_\alpha$, $\bar{\psi}_\alpha$ with $\alpha=1\ldots N$ are Grassman variables.
The partition function is
$
Z(H,\Gamma)=\int \mathcal{D}[\bar{\psi},\psi]\exp\left[-\int_0^\beta d\tau\int d^D\mathbf{x} S(H,\Gamma)\right]
$,
$\beta$ is the inverse temperature, $D$ the dimensionality, and $k_B=\hbar=1$.

We consider general $H$ and $\Gamma$ to derive WT identities,
but the physical system with individually conserved components has
$
H_{\alpha\beta}
=\mu_\alpha\delta_{\alpha\beta}$,
$
\Gamma_{\alpha\beta\gamma\delta}
=
\frac{1}{4}\left(\lambda_{\alpha\gamma}+\lambda_{\beta\delta}\right)\delta_{\alpha\delta}\delta_{\beta\gamma}
-\left(\gamma\leftrightarrow\delta\right)
$
contributing $\mu_\alpha n_\alpha$ and $\lambda_{\alpha\beta} n_\alpha n_\beta$ 
to the action where $n_\alpha=\bar{\psi}_\alpha\psi_\alpha$.  Here $\mu_\alpha$
is a chemical potential while $\lambda_{\alpha\beta}$ describes interactions 
between fermionic densities.

The action is $U(N)$ symmetric for $\mu_\alpha=\mu$ and $\lambda_{\alpha\beta}=\lambda$.
This group acts as
\begin{equation}
\label{eq:psi_un}
\psi_\alpha\rightarrow \psi'_\alpha=\sum_{\beta}U_{\alpha\beta} \psi_\beta,\
\bar{\psi}_\alpha\rightarrow \bar{\psi}'_\alpha=\sum_{\beta}\bar{\psi}_\beta U^\dagger_{\beta\alpha}
\end{equation}
with $U_{\alpha\beta}$ a unitary matrix.  Anisotropies 
in $\mu_\alpha$ and $\lambda_{\alpha\beta}$ explicitly break 
$U(N)\rightarrow U(1)^N$ describing
$N$ conserved densities in the normal state.  The superfluid state
spontaneously breaks $U(N)\rightarrow U(1)^N\rightarrow U(1)^{N-P}$ where $P$ is the number
of non-zero pairing amplitudes.

When Eq. \ref{eq:psi_un} is viewed as a field redefinition
transformation, the coupling constants also transform 
$
H_{\alpha\beta}
\rightarrow H'_{\alpha\beta}=\sum_{\gamma\delta}U_{\alpha\gamma}H_{\gamma\delta}U^\dagger_{\delta\beta}$,
$
\Gamma_{\alpha\beta\gamma\delta}
\rightarrow\Gamma'_{\alpha\beta\gamma\delta}=
\sum_{\mu\nu\rho\sigma}U_{\alpha\mu}U_{\beta\nu}\Gamma_{\mu\nu\rho\sigma}U^\dagger_{\rho\gamma}U^\dagger_{\sigma\delta}
$
in order to describe the same physical system.
By definition, $Z(H,\Gamma)=Z(H',\Gamma')$ is invariant.

This invariance arises from a field redefinition and is not a physical symmetry.
However, it still gives a WT identity (see Ref. \cite{zinn-justin-qft}) expressing 
$Z(H,\Gamma)=Z(H',\Gamma')$ under an infinitesimal transformation to first order
\begin{equation}
\label{eq:fermion_wt_identity}
\left(\mu_\alpha-\mu_\beta\right)\frac{\delta Z(H,\Gamma)}{\delta H_{\beta\alpha}}+
\sum_\gamma\left(\lambda_{\alpha\gamma}-\lambda_{\gamma\beta}\right)\frac{\delta Z(H,\Gamma)}{\delta \Gamma_{\gamma\beta\alpha\gamma}}=0
\end{equation}
for arbitrary $\alpha$, $\beta$.
Equivalently, the above equation constrains the expectation value of a redundant
operator \cite{wegner-redundant} associated with the field redefinition of Eq. \ref{eq:psi_un} 
to be zero.
This constraint is most useful in the presence of anisotropies and is
trivially satisfied in the symmetric limit.  Analyzing infinitesimal transformations
to second order may yield information about the symmetric limit.
We classify possible mean-field states by analyzing the above constraint
within the saddle-point approximation discussed next.

Introducing 
bosonic pairing $\Delta_{\alpha\beta}$ and magnetization
$M_{\alpha\beta}$ fields decouples the fermionic interaction through
a Hubbard-Stratonovich transformation 
\cite{negele-orland}.
Mean-field states are given by zero-momentum, $\tau$-independent expectation
values for the order parameters $M_{\alpha\beta}$, 
$\Delta_{\alpha\beta}$ 
satisfying the saddle-point equations
\begin{equation}
\label{eq:sp_equations}
\begin{split}
M_{\alpha\beta}=T g_M\sum_{\omega_n,\mathbf{k}}G_{\alpha\beta}
,\
\Delta_{\alpha\beta}=\frac{T g_{\Delta,\alpha\beta}}{2}\sum_{\omega_n,\mathbf{k}}
F_{\alpha\beta}
\end{split}
\end{equation}
where $\omega_n=(2n+1)\pi/\beta$ are Matsubara frequencies, $\mathbf{k}$ are momenta,
$T$ is the temperature and 
$
\lambda_{\alpha\beta}=\frac{1}{4}g_{\Delta,\alpha\beta}-\frac{1}{2}g_M\left(1+\frac{1}{N}\right)
$
for decoupling interaction anisotropies in the pairing channel.
Normal $G_{\alpha\beta}\left(i\omega_n,\mathbf{k}\right)$ 
and anomalous $F_{\alpha\beta}\left(i\omega_n,\mathbf{k}\right)$ Green's functions satisfy
\begin{equation}
\begin{bmatrix}
G&F^\dagger\\
F&-G^\dagger
\end{bmatrix}
=
\begin{bmatrix}
-\xi-H+M&\Delta^\dagger\\
\Delta&\xi^\dagger+H^\dagger-M^\dagger
\end{bmatrix}^{-1}
\end{equation}
where $\xi_{\alpha\beta}=(i\omega_n-\mathbf{k}^2/2m)\delta_{\alpha\beta}$ in a matrix notation
with suppressed indices.  As matrices, $M^\dagger=M$ and $H^\dagger=H$ are Hermitian 
while $\Delta^T=-\Delta$ is skew-symmetric.

Diagonalizing these matrices gives 
\begin{equation}
M=\sum_{i=1}^N M_i \mathbf{u}_{i} \mathbf{u}^\dagger_{i},\ 
\Delta=\sum_{i=1}^{\lfloor N/2\rfloor} \Delta_i
\left(
\mathbf{v}_{2i-1} \mathbf{v}_{2i}^T-\mathbf{v}_{2i}\mathbf{v}_{2i-1}^T
\right)
\end{equation}
where $\lfloor x \rfloor$ is the floor function, $M_i$ ($\Delta_i$) real (complex) eigenvalues,
and $\mathbf{u}_i^\dagger \mathbf{u}_j=\mathbf{v}_i^\dagger \mathbf{v}_j=\delta_{ij}$ orthonormal
eigenvectors.
From $M_{\alpha\beta} \sim \langle \bar{\psi}_\alpha \psi_\beta\rangle$, the physical interpretation of 
$\mathbf{u}_i$ is the linear combination of fermions giving magnetization $M_i$.
Similarly, $\Delta_{\alpha\beta} \sim \langle \psi_\alpha \psi_\beta\rangle$ shows
$\mathbf{v}_{2i-1}$, $\mathbf{v}_{2i}$ give the two linear combinations of fermions paired
with amplitude $\Delta_i$.

From Eq. \ref{eq:fermion_wt_identity}, the WT identity relating $M$ and $\Delta$
\begin{equation}
\label{eq:boson_wt_identity}
\left(\mu_\alpha-\mu_\beta\right)g_M^{-1} M_{\alpha\beta}+
\sum_\gamma\left(g_{\Delta,\alpha\gamma}^{-1}-g_{\Delta,\gamma\beta}^{-1}\right)
\Delta_{\alpha\gamma}\Delta^\dagger_{\gamma\beta}=0
\end{equation}
only constrains the off-diagonal elements as a matrix equation.  
In the normal state $\Delta_{\alpha\beta}=0$ implying
$M_{\alpha\beta}=0$ for $\alpha\ne\beta$.  Physically, $U(1)^N$
symmetry of $N$ conserved densities prohibits mixing between different components.

Superfluidity spontaneously breaks factors of $U(1)$ through off-diagonal elements 
of $\Delta_{\alpha\beta}$.  We now show the WT identity requires the eigenvectors 
to be of the form 
\begin{equation}
\mathbf{u}_{i,\alpha}=\mathbf{v}_{i,\alpha}=S_{i,\alpha}
\end{equation}
where $S_{i,\alpha}$ is a $N\times N$ permutation matrix with exactly one non-zero
matrix element in each row and column.  
We denote these states the DPS. 
Consider the generic case when 
$\mu_\alpha\ne\mu_\beta$, $g^{-1}_{\Delta,\alpha\beta}\ne g^{-1}_{\Delta,\gamma\delta}$.
For $N=3$, the DPS give \textit{all possible mean-field states}.  
States not of DPS form do not satisfy Eq. \ref{eq:boson_wt_identity}.
For $N>3$, the DPS give \textit{generic mean-field states}.
States not of DPS form can in principle satisfy Eq. \ref{eq:boson_wt_identity},
but off-diagonal elements are generally overdetermined by the 
saddle-point equations implying non-zero values are inconsistent.

We now discuss some properties of DPS.
They simplify Eq. \ref{eq:sp_equations} 
by decoupling them to $\lfloor N/2\rfloor$ saddle-point equations 
each involving only $\Delta_i$, $M_{2i-1}$, and $M_{2i}$.
$M$ is also diagonal and respects $U(1)^N$ symmetry as in the normal state
while $\Delta$ breaks $P$ factors of $U(1)$ where $P$ is the number of non-zero
$\Delta_i$.  The microscopic pairing wavefunction for the DPS is given by
using a Bardeen-Cooper-Schrieffer (BCS) $s$-wave state to 
pair two and only two components at a time in all possible ways.
The WT identity prohibits states given by linear combinations of the DPS. 
This gives a discrete set of mean-field states
with  $N!/P!(N-2P)!2^P$ DPS for fixed $N$ and $P$ 
giving a total of $(i/\sqrt{2})^N H_N(-i/\sqrt{2})$ distinct 
states for fixed $N$ with $H_n(x)$ the Hermite polynomial.

Quasiparticle excitations are gapped for paired components and are gapless
for unpaired components which remain a Fermi liquid.
There are at most $\lfloor N/2\rfloor$ non-zero $\Delta_i$ implying
one component is always gapless for $N$ odd.
For $N$ even, a fully gapped quasiparticle spectrum occurs only when
all $\Delta_i$ are non-zero.
Refs. \cite{hofstetter-04a,hofstetter-04b} obtained similar results for $N=3$.

For simplicity,
we construct phase diagrams in the weak-coupling or BCS regime near the 
superfluid transition for small anisotropies but
stress the preceding analysis of the DPS is general.
We minimize the free energy
\begin{equation}
\begin{split}
F=
&\Tr\left[\frac{1-2\hat{g}_Ma}{2\hat{g}_M}\delta \hat{M}\delta \hat{M}+
b\frac{T-T_c^{SYM}}{T_c^{SYM}}\hat{\Delta}^\dagger\hat{\Delta}\right]+\\
&\Tr\left[c\delta \hat{M}\hat{\Delta}\hat{\Delta}^\dagger +d \hat{\Delta}^\dagger\hat{\Delta}\hat{\Delta}^\dagger\hat{\Delta}\right]+
\sum_{\alpha} 2a\delta\hat{\mu}_\alpha\delta\hat{M}_{\alpha\alpha}+\\
&\sum_{\alpha\beta}\left(\delta\hat{g}_{\Delta,\alpha\beta}^{-1}-c\frac{\delta\hat{\mu}_\alpha+\delta\hat{\mu}_\beta}{2}\right)
\hat{\Delta}^\dagger_{\alpha\beta}\hat{\Delta}_{\beta\alpha}
\end{split}
\end{equation}
given by $F=-T\log Z$ where we take
$\mu_\alpha=E_F+\delta\mu_{\alpha}$, $g^{-1}_{\Delta,\alpha\beta}=g^{-1}_{\Delta}+\delta g^{-1}_{\Delta,\alpha\beta}$,
$M_{\alpha\beta}=g_M n(E_F)+\delta M_{\alpha\beta}$ with $E_F$ the Fermi energy, $n(E_F)$ the 
free fermion density and from here on, quantities
with a hat are rescaled with respect to $T_c^{SYM}$, the critical temperature
without anisotropies.
Here $a,b,c,d$ are Ginzburg-Landau parameters with $a,b,d\sim \hat{E}_F^{(D-2)/2}$ 
proportional to the density of states
at $\hat{E}_F$ describing particle-hole symmetric contributions.
In contrast, $c\sim \hat{E}_F^{(D-4)/2}$ for $D=2$ and $c\sim \hat{E}_F^{(D-4)/2}\log \hat{E}_F$
for $D\ne 2$ is essentially proportional to the derivative of the density of states at $\hat{E}_F$ and 
describes particle-hole symmetry breaking contributions.

$U(N)$ symmetric terms are under $\Tr$, the matrix trace.
Notice the $\delta \hat{M}\hat{\Delta}\hat{\Delta}^\dagger$ term
where $\hat{\Delta}\hat{\Delta}^\dagger$ acts as an external field to $\delta\hat{M}$.
On group theoretical grounds, this term is non-vanishing only for $N\ge 3$.
Terms outside the trace explicitly break $U(N)\rightarrow U(1)^N$, including
a term quadratic in $\hat{\Delta}$.    
The structure of global phase diagrams follows from these two terms.

Pairing $\hat{\Delta}\hat{\Delta}^\dagger$ drives magnetization 
$\delta\hat{M}$ unless $\hat{\Delta}\hat{\Delta}^\dagger\propto \mathbf{1}$
is particle-hole symmetric with $\mathbf{1}$ the identity matrix.
In this case, a shift in chemical potentials described by $\hat{M}$
yields no gain in condensation energy.
Only when all components pair and $|\Delta_i|=|\Delta_j|$ does this occur.  
For $N$ odd, one component is always unpaired so magnetization always develops.
For $N$ even, only when the $N/2$ independent equations determining $\Delta_i$
give $|\Delta_i|=|\Delta_j|$ is it possible to have pairing without magnetization.

A similar situation occurs for $p$-wave pairing in $^3$He or the organic superconductors.
Unitary states describe pairing decoupled from magnetization while
non-unitary states describe pairing coupled to magnetization \cite{leggett-75}.
Only a single constraint $|\mathbf{d}\times \mathbf{d}^*|=0$ on the $\mathbf{d}$-vector
describing $p$-wave pairing is necessary for a unitary state.  
It is more difficult for the $N/2$ independent equations
determining $\Delta_i$ to give $|\Delta_i|=|\Delta_j|$ for the analog of unitary states
in multicomponent $s$-wave pairing.

We now discuss the $N=3$ phase diagram in Fig. \ref{fig:n3_phasediagram}
with given interactions satisfying the generic condition
$\delta\hat{g}^{-1}_{\Delta,\alpha\beta}\ne \delta\hat{g}^{-1}_{\Delta,\gamma\delta}$.  We consider
both fixed chemical potential $\hat{\mu}_\alpha$ and fixed particle density $n_\alpha$ as
\begin{equation}
\begin{split}
\hat{\mu}_\alpha&=\hat{E}_F \mathbf{x}_{0,\alpha}+\hat{\mu}_1 \mathbf{x}_{1,\alpha}+\hat{\mu}_2 \mathbf{x}_{2,\alpha},\\
n_\alpha&=n(E_F) \mathbf{x}_{0,\alpha}+n_1 \mathbf{x}_{1,\alpha}+n_2 \mathbf{x}_{2,\alpha}
\end{split}
\end{equation}
where $\mathbf{x}_0=[1,1,1]$, $\mathbf{x}_1=[1,-1,0]/\sqrt{2}$, $\mathbf{x}_2=[1,1,-2]/\sqrt{6}$
and $\hat{\mu}_i$, $n_i$ parameterize anisotropies.

First consider fixed $\hat{\mu}_\alpha$ (top Fig. \ref{fig:n3_phasediagram}). 
When $T>T_c^{SYM}$, small anisotropy favors the normal state.
Increasing anisotropy favors pairing by increasing the density of 
states for some components at the expense of others.  
This drives a second order
transition into one of the three DPS.  Tuning the direction
of the anisotropy drives first order transitions between DPS
when two of these states are degenerate along lines of enhanced symmetry.
First order lines terminate at bicritical points from
which metastability regions where an additional DPS is locally stable
branch out.  This is the expected behavior for
quadratic symmetry breaking \cite{nelson-76,amit-77}.
Bicritical points terminate at the $U(3)$ symmetric multicritical point
when $T=T_c^{SYM}$.
Below $T_c^{SYM}$, first order lines separate DPS and a region near small anisotropy
where all three DPS are (meta)stable appears.

The Maxwell construction gives fixed $n_\alpha$ phase diagrams (bottom Fig. \ref{fig:n3_phasediagram})
from those for fixed $\hat{\mu}_\alpha$.
The compressibility tensor 
$\kappa_{\alpha\beta}\propto-\partial^2 F/\partial \hat{\mu}_\alpha\partial\hat{\mu}_\beta$ 
at fixed $\hat{\mu}_\alpha$ has positive eigenvalues except at boundaries between DPS.
Here, $\hat{\Delta}\hat{\Delta}^\dagger$ and thus $\delta\hat{M}$ jump discontinuously. 
First order lines at fixed $\hat{\mu}_\alpha$ when two DPS are degenerate 
expand into phase-separated mixtures of those two states at fixed $n_\alpha$.  
For $T<T_c^{SYM}$, the zero anisotropy point where all three DPS are degenerate expands
into a phase-separated mixture of those three states at fixed $n_\alpha$. 

Now consider the $N=4$ phase diagram in Fig. \ref{fig:n4_phasediagram}
with fixed chemical potential given by
\begin{equation}
\begin{split}
\hat{\mu}_\alpha=&
\hat{E}_F \mathbf{y}_{0,\alpha}+
\hat{\mu}_0\cos(\theta)\mathbf{y}_{1,\alpha}+\\
&\hat{\mu}_0\sin(\theta)\cos(\phi)\mathbf{y}_{2,\alpha}+
\hat{\mu}_0\sin(\theta)\sin(\phi)\mathbf{y}_{3,\alpha}
\end{split}
\end{equation}
where $\mathbf{y}_0=[1,1,1,1]$, $\mathbf{y}_1=[1,-1,0,0]/\sqrt{2}$, $\mathbf{y}_2=[1,1,-2,0]/\sqrt{6}$,
$\mathbf{y}_3=[1,1,1,-3]/\sqrt{12}$ and $\theta$, $\phi$ 0 the anisotropies.

For $N>3$, condensation energy favors development of 
more pairing amplitudes as $T$ is lowered.
The normal state competes with the six DPS with
only one pairing amplitude for $T>T_c^{SYM}$ (top right Fig. \ref{fig:n4_phasediagram}).
However, the three DPS with two pairing amplitudes
and all components paired eventually dominate the phase diagram for $T<T_c^{SYM}$
(top middle Fig. \ref{fig:n4_phasediagram}).
Phase diagrams still exhibit
second order transitions from the normal state to DPS and
first order transitions between DPS.

We now comment on trapping effects and detection methods for applications to ultracold
atoms.  In the large particle number BCS regime, the local 
density approximation accurately maps phase diagrams at fixed chemical potential
to phase diagrams with trapping.
DPS are distinguished by both densities and pairing amplitudes for different components.
State-selective imaging of density distributions \cite{mit-06,rice-06} and
radio-frequency spectroscopy of the pairing gap \cite{chin-04} can be used to detect signatures of 
the various paired states.

In summary, we have studied the essential role magnetism plays in superfluidity of multicomponent fermions.
By analyzing constraints imposed by Ward-Takahashi identities, we classified the allowed
mean-field pairing states with both magnetization and pairing order parameters
and used them to construct global phase diagrams.  These phase diagrams
have a rich structure with first and second order transitions meeting at multicritical points as
well as metastability and phase separated regions.  We discussed applications 
to ultracold fermions.

We acknowledge useful discussions with Walter Hofstetter.
This work was supported by NSF grant DMR-0132874, Harvard-MIT CUA, AFOSR, and NDSEG.

\bibliography{references}

\end{document}